\documentclass[useAMS,usenatbib]{mn2e}
\usepackage[dvips]{graphicx}
\usepackage{amssymb}

\title[X-ray absorption and occultation in LS 5039]{X-ray absorption and occultation in LS 5039}

\author[A. Szostek and G. Dubus]{
Anna Szostek$^{1,2}$\thanks{E-mail: aszostek@obs.ujf-grenoble.fr} and Guillaume Dubus$^{1}$\\
$^{1}$Laboratoire d'Astrophysique de Grenoble, UMR 5571 Universit\'e Joseph Fourier Grenoble I / CNRS, BP 53, 38041 Grenoble, France\\
$^{2}$Astronomical Observatory, Jagiellonian University, Orla 171, 30-244 Krak\'ow, Poland}

\date{Accepted September 2010}

\pagerange{\pageref{firstpage}--\pageref{lastpage}}\pubyear{2009}

\begin{document}

\maketitle

\label{firstpage}

\begin{abstract}
Gamma-ray binaries are systems containing a massive star and a compact object that have been detected up to TeV energies. The high energy emission could result from particle acceleration in the region where the stellar wind from the massive star interacts with the relativistic wind from a young pulsar. LS 5039 has the most compact orbit amongst gamma-ray binaries and its X-ray lightcurve shows a stable modulation synchronized with the orbital period. Photoelectric absorption of X-rays in the O star wind and occultation of the X-ray emitting region by the massive star can alter the X-ray lightcurve and spectrum along the orbit. Yet, the X-ray spectrum and lightcurve of LS 5039 do not show intrinsic absorption or X-ray eclipses. We study these effects in the framework of the pulsar wind scenario as a function of the binary inclination angle, the stellar wind mass-loss rate and the size of the X-ray emitter. An extended X-ray emission region $\ga 3 R_\star$ appears necessary to reconcile the pulsar wind scenario with observations.

\end{abstract}


% =========================================================
\section{Introduction}

Gamma-ray binaries are systems containing a massive star and a compact object that emit most of their power at energies above 100 MeV. They have been detected up to very high energy (VHE) gamma-rays and are thus sites of particle acceleration up to multi-TeV energies. There are three established gamma-ray binaries PSR B1259-63 \citep{hesspsrb1259}, LS 5039 \citep{hessls5039,fermils5039} and LS I+61 303 \citep{magiclsi61303,fermilsi61303} and a candidate binary HESS J0632+057 \citep{discoveryhessj0632}. 

The VHE emission in gamma-ray binaries is thought to be due to Compton upscattering of UV photons from the massive star by energetic electrons. Electron acceleration could take place either in a relativistic jet (microquasar scenario, \citealt*{romero2005}; \citealt*{paredes2006}; \citealt{dermer2006}) or in the shocked wind of a young pulsar (pulsar wind scenario, \citealt{maraschi1981,dubus2006}), where the shock results from the interaction between the pulsar wind and the stellar wind of the companion star. The latter model is known to be operating in case of PSR B1259-63 (\citealt*{tavani1994}; \citealt*{Kirk1999}). The nature of the compact object and the electron acceleration site are uncertain in the remaining gamma-ray binaries.

Gamma-ray binaries have also been observed in X-rays where their properties differ from those of high-mass X-ray binaries (HMXB). The X-ray spectrum of gamma-ray binaries are power-laws but show no apparent cutoffs up to hundreds of keV. The X-rays are thought to be due to non-thermal synchrotron or inverse Compton emission. Gamma-ray binaries also do not display X-ray outbursts and state transitions as usually seen in accreting binaries.

LS 5039 is the most compact gamma-ray binary, composed of an unknown compact object in a 3.9 day orbit around a  O6.5V star \citep{casares2005}. It has a regular behavior in both gamma-rays and X-rays. The X-ray lightcurve shows an orbital modulation with a remarkable long-term stability \citep{kishishita2009,takahashi2009,hoffmann2009}. The modulation is present at low (1--10 keV) and medium (10--40 keV) X-ray energies, with minimum and maximum flux at superior and inferior conjunction respectively. The location of the minima and maxima suggest the modulation is a geometrical effect related to the orientation of the binary with respect to the observer, rather than due to physical changes taking place in the shocked winds region when the compact object travels on its elliptical orbit.

X-ray absorption in the stellar wind and occultation of the X-ray emitting region by the massive star both result in orbital modulations with the correct phases for flux minimum and maximum. However, for LS 5039 there is no evidence for an absorption excess due to the stellar wind at any orbital phase, even at superior conjunction \citep{takahashi2009}. This has been used to argue that the X-ray emitting region must be far out from the system, where  the column density of material crossed by the X-rays is small \citep{bosch2007,bosch2009}. Here, we re-examine this question in detail, taking into account the 3D geometry of the interaction in the pulsar wind scenario to compute model lightcurves. This is used to derive constraints on the binary parameters from the lack of absorption and occultation signatures in the observed lightcurve.

The outline of this article is as follows. The model is described in \S\ref{s:model} and applied in \S\ref{s:absorption} to estimate the size of the X-ray emitting region based on the observed hydrogen column density. The  occultation of the X-ray emitting region by the massive star is studied in section \S\ref{s:occultation}. Finally, \S\ref{s:discussion} discusses the findings and presents the conclusions.

\begin{figure}
\centerline{\includegraphics[width=75mm]{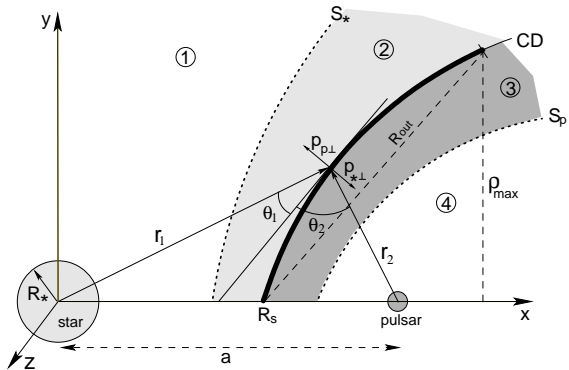}}
\caption{The schematic 2D illustration of the model where the stellar wind and pulsar wind collide. The sizes of the pulsar and star are not to scale.} 
\label{shock_schematics}
\end{figure}

\section{The geometric model}
\label{s:model}

\subsection{Shape of the shock}
In the pulsar wind scenario, X-rays are emitted by particles accelerated in the shock region where the pulsar wind and the stellar wind of the massive star collide (Fig. \ref{shock_schematics}). The interaction region is bounded by two termination shocks $S_{\star}$ and $S_{\rm p}$. Downstream (zones 2 and 3), two shocked winds are separated by a tangential contact discontinuity (CD). Upstream (zones 1 and 4), the winds behave as in the case of single star/pulsar. In this example, the stellar wind momentum dominates over the pulsar wind.
The stagnation point $R_s$ between the two winds along the line joining the stars is found by equating the ram pressures from the two winds:
\begin{equation}
p_{\star}(R_s) = {\dot M v(R_s) \over 4 \pi R_s^2}={\dot E \over 4 \pi c  (a-R_s)^2}=p_{\rm p}(a-R_s),
\end{equation}
where $a$ is the binary orbital separation, $\dot{M}$ is the stellar wind mass loss rate and $\dot{E}$ is the pulsar spindown power. The stellar wind velocity $v$ at the radial distance $r$ from the star's centre is given by a $\beta$-velocity law 
\citep*{castor1975}
\begin{equation}
v(r) = v_{\infty} \left(1- {R_{\star} r_0 \over r} \right)^{\beta},
\label{eq:vwind}
\end{equation}
where $r_0 = 1-(v_0/v_{\infty})^{1/\beta}$ with $v_0$ the initial wind velocity, $v_{\infty}$ is the wind terminal velocity, $R_{\star}$ is the stellar radius and $\beta\approx 1$ is a parameter describing wind acceleration. 

%%%%%%%%%%%%%%%%%%%%%%%%%%%%%%%%%%%%%%%%%%%%%%
The detailed structure of the shock will depend on the momenta of the winds and on radiative cooling of the gas. Efficient cooling will tend to collapse the termination shocks onto the CD. The shock structure can also be affected by mixing instabilities at the interface and by orbital motion \citep*[e.g.][]{stevens1992}. There is no general semi-analytic description of the shock structure but a description of the CD can be obtained. The CD is defined as the surface where the perpendicular components of the ram pressures balance each other  $p_{\star \perp} = p_{\rm p \perp}$. For each point in space one can then define a dimensionless parameter  
\begin{equation}
\eta(r_1) \equiv  {\dot E \over \dot M v(r_1) c } = {r_2^2 \sin^2\theta_1 \over r_1^2 \sin^2\theta_2}, 
\label{eq:eta}
\end{equation}
where $\theta_1$ an $\theta_2$ are angles between the line tangential to the CD at the given point and the direction towards the star or pulsar respectively as illustrated in Fig. \ref{shock_schematics}. $r_1$ and $r_2$ are distances from the star and pulsar respectively. The value of $\eta$ at the stagnation point ($\theta_1 = \theta_2 = \pi/2$) equals
\begin{displaymath}
 \eta_0 = \eta(R_{\rm s}) =  {\dot E \over \dot M v(R_{\rm s}) c } =  {(a-R_{\rm s})^2 \over R_{\rm s}^2} =
 \label{eta0}
 \end{displaymath}
 \begin{equation}
 = 0.05 \left({ \dot E \over 10^{36}\ {\rm erg}\ {\rm s}^{-1}  }\right)\left( {\dot M \over 10^{-7}\ {\rm M}_{\odot}\ {\rm yr}^{-1}  }\right)^{-1} \left [ {v(R_{\rm s}) \over 10^8\ {\rm cm\ s}^{-1}} \right]^{-1}.
 \label{eq:eta0}
 \end{equation}
The $\eta_0$ depends on orbital phase via $v(R_{\rm s})$ and parametrizes the shape of the shock, which is then obtained by solving the differential equation (Eq. 6 in \citealt*{antokhin2004})
\begin{equation}
{dx \over dy} = {1 \over y} \left \{ x- 
{a r_1^2 (x,y) \sqrt{\eta\left [r_1(x,y) \right]}  \over   r_1^2(x,y) \sqrt{ \eta\left [r_1(x,y)\right ] } + r_2^2(x,y)}        \right \},
\label{eq:antokhin}
\end{equation}
with the initial condition $x = R_{\rm s}$ and $y=0$. Numerical simulations show that this holds reasonably well in the case of the interaction with a relativistic pulsar wind \citep{bogovalov2008}. The solution is a 2D profile of the CD $x=f(y)$. The CD is symmetrical with respect to $x$-axis, thus it is best represented in cylindrical coordinates $(\rho, \psi,x)$ as $x=f(\rho)$, where $\rho=\sqrt{y^2+z^2}$ and $\psi=\arctan(z/y)$. 

\begin{figure}
\centerline{\includegraphics[width=75mm]{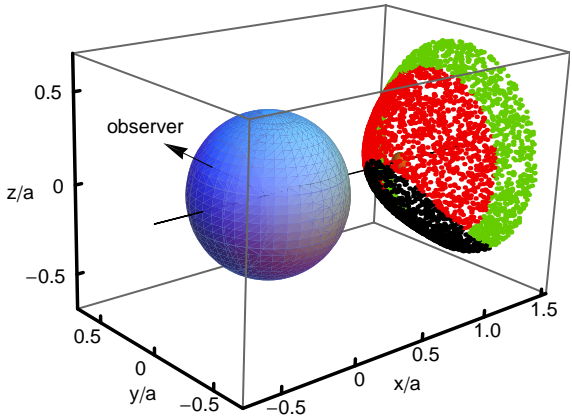}}
\caption{3D illustration of LS 5039 in the framework of the pulsar wind scenario. The plot is in units of orbital separation and the massive star is plotted to scale. The dark sphere is the massive star, the light grey (green) points represent the CD surface, the dark grey (red) points are the points that emit X-rays, whereas black points are in the shadow of the massive star. The arrow represents the direction to the observer at inclination $60^{\circ}$ and at the phase of superior conjunction. } 
\label{3Dshock_schematics}
\end{figure}

An example 3D representation of the model is shown in Fig.~\ref{3Dshock_schematics}. In cases where $\eta_0 < 1$ the stellar wind dominates over the pulsar wind. When $\eta_0 \ll 1$, the opening angle of the CD $\alpha \lesssim 45^{\circ}$ (where $\alpha$ is an angle between $x$-axis and a line tangential to the CD at a large distance from the stagnation point) and the CD wraps around the pulsar creating a tail-like structure. For $\eta_0 >1$ the opening angle $\alpha > 90^{\circ}$ and the CD curves around the massive star. Note that there is a maximum value of $\eta_0$ associated with a minimum realizable distance $R_{\rm s}^{\rm min}$ between the star and CD. This is approximately the place where $p_{\star}(r)$ has its maximum. Stable balance is lost if the pulsar wind moves beyond this point:  then the pulsar wind overwhelms the stellar wind and is stopped at the star surface.

The shape of the CD will be correct up to the point where the Coriolis force curves the shock structure. Assuming that the stellar wind is collimating the pulsar wind, this happens at a distance $\sim v_\infty P_{\rm orb}$, or about 5.4 a.u. (130 R$_\star$) for LS 5039 with the parameters given in \S\ref{s:ls5039}. Closer in, the orientation of the surface can also be altered by orbital motion: the skew angle $s$ is given by $\tan s =v_{\rm orb}/v(a)$ where $v_{\rm orb}$ is the pulsar orbital velocity.
%
%%%%%%%%%%%%%%%%%%%%%%%%%%%%%%%%%%%%%%%%%%%%%%
\subsection{X-ray emission}
\label{ss:xray}

In the pulsar wind scenario, the X-ray emission is primarily due to synchrotron emission in the shocked pulsar wind region (zone 3 in Fig. \ref{shock_schematics}) which is tenuous and optically thin to X-rays. The stellar wind region may also be a source of X-rays, either because of the kinetic energy dissipated in the colliding winds (zone 2) or due to small-scale instabilities in the stellar wind (zone 1) \citep*{puls2008}. In both cases the emission is softer ($kT\approx 0.5$ keV) and weaker ($10^{30}$--$10^{33}$  erg\ s$^{-1}$, e.g. \citealt{stevens1992})  than the observed X-ray emission from LS 5039 ($\approx 10^{33}$  erg\ s$^{-1}$, \citealt{takahashi2009}). Detecting X-ray line emission would provide valuable diagnostics of the stellar wind but this emission is most likely swamped by synchrotron emission. 

The cooling parameter $\chi\equiv t_{\rm cool}/t_{\rm dyn}$ of the O star wind \citep{stevens1992} is $\approx 1$ in LS 5039, at the limit of efficient radiative cooling. Isothermal (radiatively efficient) shocks have small widths. The pressure in the shocked pulsar wind will mostly be due to the lowest-energy electrons if the particles have a steep distribution. These electrons have long cooling times and the shock may be more closely approximated as adiabatic rather than isothermal. Complex numerical simulations are required to obtain the detailed structure and emissivity of the shock region. At this stage, we assume that the shocked regions 2 and 3 have a negligible width and that the emissivity is uniform over the 2D shock surface, which should suffice to capture the geometrical effects that we want to investigate. 

The 2D shock surface is approximated by the CD surface for $\rho \in \langle 0, \rho_{\rm max} \rangle$ and zero otherwise, with $\rho_{\rm max}$ the cylindrical radius at which a sphere centered at $(R_{\rm s},0,0)$ of radius $R_{\rm out}$ intersects the CD (see Fig. \ref{shock_schematics}). The size of the emitting region is conveniently parametrized by $R_{\rm out}$ in the following. The total flux from the emitting volume is
\begin{equation}
F(E) = \int_V j_{\rm E} \ dV = \int_S \lambda\ j_{\rm E} \ dS,
\end{equation}
where $j_{\rm E}$ is the unit volume emissivity, and $\lambda j_{\rm E}$ is the constant surface emissivity. 
The unit surface element of CD equals 
\begin{equation}
dS =  \sqrt {1+ \left ( {\partial f \over \partial \rho} \right )^2}\ \rho\ d\rho\ d\psi.
\label{eq:ds}
\end{equation}

\subsection{Absorption and occultation}
The uniform emission of X-rays is affected in two, orbital phase-dependent, ways. First, X-rays can undergo photoelectric absorption as they cross the dense stellar wind. The wind density at the distance $r$ from the stellar center is
\begin{equation}
n(r) = {\dot M \over 4 \pi \mu m_{\rm H} r^2 v(r)},
\end{equation}
where $m_{\rm H}$ is hydrogen mass while $\mu=1.3$ is the mean molecular weight. Second, at some phases and for some inclinations, parts of the emitting surface of the CD are occulted by the massive star and are thus invisible to the observer (black regions in Fig. \ref{3Dshock_schematics}). 

In order to estimate the impact of absorption and occultation on the observed flux, a random sample of $N$ points is uniformly distributed on the surface of CD out to $R_{\rm out}$ (see Fig. \ref{3Dshock_schematics} and Appendix A). Each point corresponds to a $N$-th part of the emitting surface and has a flux equal $1/N$ (the total flux is normalized to unity). The observed flux at given energy $E$, inclination angle $i$ and orbital phase $\phi$ is 
\begin{equation}
F(E,i,\phi) = {1 \over N} \sum_{j=1}^N \exp[-\sigma(E) N_{{\rm H}j}(i,\phi)]\  \zeta_j(i,\phi),
\label{eq:flux}
\end{equation}
where $N_{{\rm H}j}$ is the stellar wind's hydrogen column density obtained from integration of $n(r)$ along the line of sight from point $j$, $\sigma$ is the photoelectric cross section of the plasma which is a function of photon energy $E$. If the plasma is ionized, $\sigma$ also depends on plasma ionization parameter which varies along the line of sight. $\zeta_i$ is the occultation function which value is 0 whenever a line of sight crosses the interior of the massive star and 1 otherwise.  

In calculations of $N_{\rm H}$ the parts of the line of sight that cross the pulsar side of the CD (zones 3 and 4 in Fig. \ref{shock_schematics}) are considered to be empty since neither unshocked nor shocked pulsar wind have sufficient densities to absorb X-rays. For simplicity, we also assume that zone 1+2 are described by Eq.~\ref{eq:vwind} all the way from the stellar surface out to the CD {\it i.e.} we ignore the density enhancement in the shocked stellar wind (zone 2). Again (\S\ref{ss:xray}), detailed numerical simulations would be required to model this appropriately. An estimate can be derived using the thin shell colliding wind model of \citet*{canto1996}. The surface density of the shock is $\approx 3\dot{M}\sqrt{\eta}/(16\pi a v)$ in the orbital plane, where it is maximum.  This corresponds to a column density across the shock $N_{\rm H, shock}\approx 7\ 10^{19} \dot{M}_{10^{-7} {\rm M}_\odot{\rm /yr}} \eta^{1/2}_{0.01} a^{-1}_{\rm 0.1\ au} v_{\rm 2000\ km/s}^{-1} {\rm cm}^{-2}$, compared to the wind column density $N_H\approx 2\ 10^{21} {\rm cm}^{-2}$ from $R_\star$ to infinity (for the same wind parameters).  $N_{\rm H, shock}$ is also about 30 times less than the inferred wind column density at superior conjunction (see \S\ref{s:absorption}). We conclude that taking into account the density enhancement in the shocked region is unlikely to change our results significantly.

%%%%%%%%%%%%%%%%%%%%%%%%%%%%%%%%
\subsection{Parameters for LS 5039}
\label{s:ls5039}

We adopt the following binary parameters for LS 5039: $v_{\infty}= 2.4 \times 10^8$ cm s$^{-1}$, $R_{\star} = 9.3 R_{\odot}$, $M_{\star} = 23 M_{\odot}$ and $T_{\star}=3.9 \times 10^4$ K. For the wind of the massive star, we adopt $\beta=1$ and $v_0 = 2 \times 10^6$ cm~s$^{-1}$ (Eq.~\ref{eq:vwind}). The value of $\dot M$ in LS 5039 is not well constrained and varies from $\sim 3 \times 10^{-8}$ to $7.5 \times 10^{-7}$ M$_{\odot}$~yr$^{-1}$, with values of several $10^{-7}$ M$_{\odot}$~yr$^{-1}$ favoured in the most recent litterature  (\citealt{kudritzki2000,mcswain2002,mcswain2004,casares2005}; \citealt*{szalai2010}). These mass loss rates are derived from H$\alpha$ line fitting, a diagnostic which is known to be affected by wind clumping \citep{puls2008}. The true $\dot{M}$ could be a factor 2--3 lower than the $\dot{M}$ estimated from H$\alpha$. The large range of $\eta$ that we explore covers this uncertainty.

The binary inclination angle is $i\ga 40^{\circ}$ if the binary contains a neutron star. The orbit eccentricity is $e=0.33$, while the angle of the line of nodes is $\omega=236^{\circ}$ \citep{aragona2009}. For these orbital parameters the periastron, apastron, pulsar superior and inferior conjunction correspond to orbital phases $\phi=$ 0, 0.5, 0.045 and 0.67 respectively. The respective binary separations are $1.4 \times 10^{12}$, $2.82 \times 10^{12}$, $1.46 \times 10^{12}$, and $2.59 \times 10^{12}$ cm.  The skew angle is $s=23^{\circ}$ at periastron and $8^{\circ}$ in apastron. This has a negligible impact on the conclusions and is not taken into account.

A strict lower limit on the value of $\dot E$ in LS 5039 is $\simeq 10^{35}$ erg s$^{-1}$ which implies 100\% radiative efficiency to gamma rays (at a distance of 2-3 kpc). At the other end of the scale, the pulsar wind will impact the star surface beyond a certain $\dot E_{\rm max}$ set by $\dot M$ and the binary separation at periastron. Such a situation is thought to occur in the black widow pulsars where the wind from the low-mass stellar companion is very weak  \citep{phinney1988}. If the pulsar wind is dominated by the kinetic energy of the cold e$^+$ e$^-$ pairs then radiative braking of the pulsar wind by inverse Compton losses might still allow a shock to form just above the O star surface \citep*{cerutti2008}. The magnetic field of the O star may also be sufficient to hold off the pulsar wind \citep{harding1990}. However, there is no observational evidence that the stellar wind in LS 5039 collapses due to the relativistic wind momentum impinging on its pulsar-facing hemisphere. \citet{mcswain2004} find no differences in UV spectra taken at two different orbital phases (0.40 and 0.63 based on ephemeris of \citealt{aragona2009}), the latter one close to inferior conjunction of the compact object where the effect of the O star wind quenching by the pulsar wind should be most visible.  Assuming the O star wind does not collapse implies a maximum $\eta_{\rm max} \simeq 0.6$, reached at periastron, for which the CD opening angle is $\approx 62^{\circ}$. For the lower and upper limit on $\dot M$, the corresponding $\dot E_{\rm max}$ varies between $\sim 1.5 \times 10^{36}$ erg s$^{-1}$ and $3.8 \times 10^{37}$ erg s$^{-1}$ respectively, well below the $\dot E$ of the Crab pulsar ($4.6 \times 10^{38}$ erg~s$^{-1}$). In the following calculations we often use the value of $\eta_0=0.004$ to examine the model predictions in the case of extreme stellar wind domination over the pulsar wind. At this value of $\eta_0$ the CD opening angle is only $\sim 10^{\circ}$. 
 
There is also no evidence for X-ray ionization of the stellar wind \citep{mcswain2004}, suggesting that the impact of ionization on X-ray absorption can be neglected. We confirmed the validity of this assumption for LS 5039 with the photoionization code XSTAR \citep{xstar}. In the simulation, an optically thin plasma cloud was illuminated by a source with a spectrum composed of a power law with spectral index 1.5 and 1--10 keV luminosity $6.03 \times 10^{33}$ erg s$^{-1}$ (\citealt{takahashi2009}, assuming a 2.5 kpc distance to LS 5039 from Earth) and a black body (stellar continuum) with a temperature $T_{\star}$, radius $R_{\star}$ and luminosity $L_{\star} = 6.9 \times 10^{38}$ erg s$^{-1}$. The measured opacities in the 1--10 keV range did not differ from those of the cold plasma described by \citet{morrison1983}. The intense UV radiation of the massive star does not influence the plasma opacity at energies above 0.5 keV.

%%%%%%%%%%%%%%%%%%%%%%%%%%%%%%%%%%%%%%%%%%%%%%
\section{X-ray absorption in LS 5039}
\label{s:absorption}

Observations show no signatures of X-ray absorption by the stellar wind in LS 5039. The {\it XMM-Newton} and {\it Suzaku} spectra of LS 5039 are absorbed by an equivalent hydrogen column density consistent with the Galactic value. The upper limit on the intrinsic column density at the phase of superior conjunction (where the line of sight column density related to the stellar wind is highest) is as low as $N_{\rm H}^{\rm max}=2.6 \times 10^{21}$ cm$^{-2}$ \citep{bosch2007,takahashi2009}.

The photoelectric absorption depends on four parameters, $\dot M$, $i$, $R_{\rm out}$ and $\eta_0$. This can be used to estimate the maximum allowed $\dot M_{\rm max}$ for which the line of sight column density does not exceed $N_{\rm H}^{\rm max}$. For example, for a point X-ray source located at the pulsar position, at the phase of superior conjunction, and at inclination angle $i = 40^{\circ}$, the upper limit on the stellar wind mass loss rate is $\dot M_{\rm max}=7.3 \times 10^{-8}$ M$_{\odot}$ yr$^{-1}$. For a point source located at the stagnation point, the $\dot M_{\rm max}$ is lower, since the line of sight probes regions of the stellar wind with higher density. For an X-ray emitting point source located at $R_{\rm s}$ and for which $\eta_0=\eta_{\rm max}$, then $\dot M_{\rm max} = 2.3 \times 10^{-8}$ M$_{\odot}$ yr$^{-1}$. 
\begin{figure}
\centerline{\includegraphics[width=75mm]{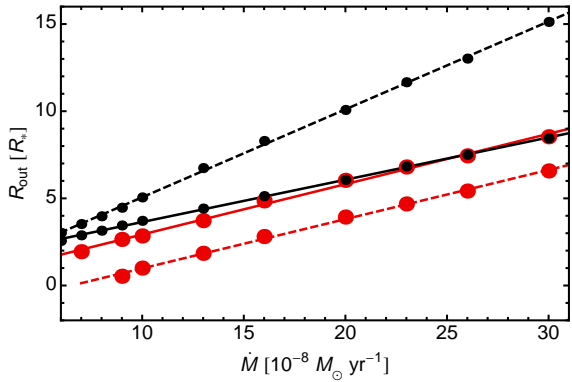}}
\caption{The minimum size of the X-ray 1 keV emission region at superior conjunction. The dots mark the results of calculations, the lines are linear fits to the points. The dashed lines correspond to $\eta_0=0.004$ at superior conjunction, the solid lines correspond to maximum value of $\eta_0 =\eta_{\rm max}=0.6$ at periastron. The grey (red) and black correspond to $i=40^{\circ}$ and $i=60^{\circ}$ respectively.} 
\label{extended_source}
\end{figure}

These values of $\dot M_{\rm max}$ are at the lower end of the $\dot M$ scale for the Galactic O stars, \citep[$\sim 10^{-8}-10^{-6}$ M$_{\odot}$ yr$^{-1}$, e.g.][]{mokiem2007}. $\dot M_{\rm max}$ could be higher if the wind was highly ionized but this is unlikely (see \S\ref{s:ls5039}).
 $\dot M_{\rm max}$ could also be higher if the X-ray emitter is away from the massive star, e.g., in the form of a jet, emitting X-rays at a distance $> 10^{12}$ cm where the stellar wind is rarified \citep{bosch2007}.  Another possibility is that the X-ray emitting region is not point-like but large compared to the size of the massive star.

For the measured value of $N_{\rm H}^{\rm max}$ and as a function of $\dot M_{\rm max}$, we estimate the minimum size of the X-ray emitting region for which intrinsic absorption would not manifest itself in the X-ray spectrum. 
The upper limit $N_{\rm H}^{\rm max}$ is based on X-ray data fitting which assumed that an entire (point or extended) X-ray source is covered by a uniform absorbing gas cloud so that
\begin{equation}
F(E) = \exp[-\sigma(E)N_{\rm H}],
\end{equation}
where the intrinsic flux equals 1. In our case however, where an extended X-ray source is absorbed by a stellar wind, the line of sight column density of each X-ray emitting surface element is different and the observed flux is given by Eq.~\ref{eq:flux}. In general, the observed spectra of a source covered by an uniform absorber and of a source covered by stellar wind are different ($\lesssim 5 $ keV). We assume that the spectral fitting is not able to distinguish between different types of absorbers if the fluxes at 1 keV are equal i.e.
\begin{equation}
{1 \over N} \sum_{j=1}^{N} \exp[-\sigma(1 {\rm keV})N_{{\rm H}j}(i,\phi)] = \exp[-\sigma(1 {\rm keV})N_{\rm H}^{\rm max}],
\label{eq:equation_size}
\end{equation}
where $\sigma({\rm 1 keV}) = 2.4 \times 10^{-22}$ cm$^2$ \citep{morrison1983}.
This is justified, because at energies $\gtrsim 2$ keV, the photoelectric cross-section decreases quickly and the difference between the two absorbed spectra is relatively small. Furthermore, the spectrum is most sensitive to absorption at energies  $\lesssim 2$ keV, but below 1 keV the sensitivity of the instruments starts to drop and the data errors increase which may also make it difficult to distinguish between different models. 

In order to obtain the lower limit on the size of the X-ray emitting region at 1 keV, we explore the parameter space ($\dot M$, $\eta$, $R_{\rm out}$, $i$) to find solutions that satisfy Eq.~\ref{eq:equation_size} at superior conjunction. In this calculation we reject points which are occulted by the star (adjusting $N$ accordingly) to test only the effect of absorption. 

The estimated minimum size of the 1 keV X-ray source is shown in Fig. \ref{extended_source}. The size of the emitting region must be larger to compensate for the denser gas when the  mass loss rate is increased. For $\eta_0 \simeq \eta_{\rm max}$ and at large emitter size, the results are not sensitive to inclination changes. The highly collimated pulsar wind shock ($\eta_0 \ll 1$) needs to have a large emitting surface to explain the observations.

\begin{figure}
\begin{tabular}{c}
\centerline{\includegraphics[width=75mm]{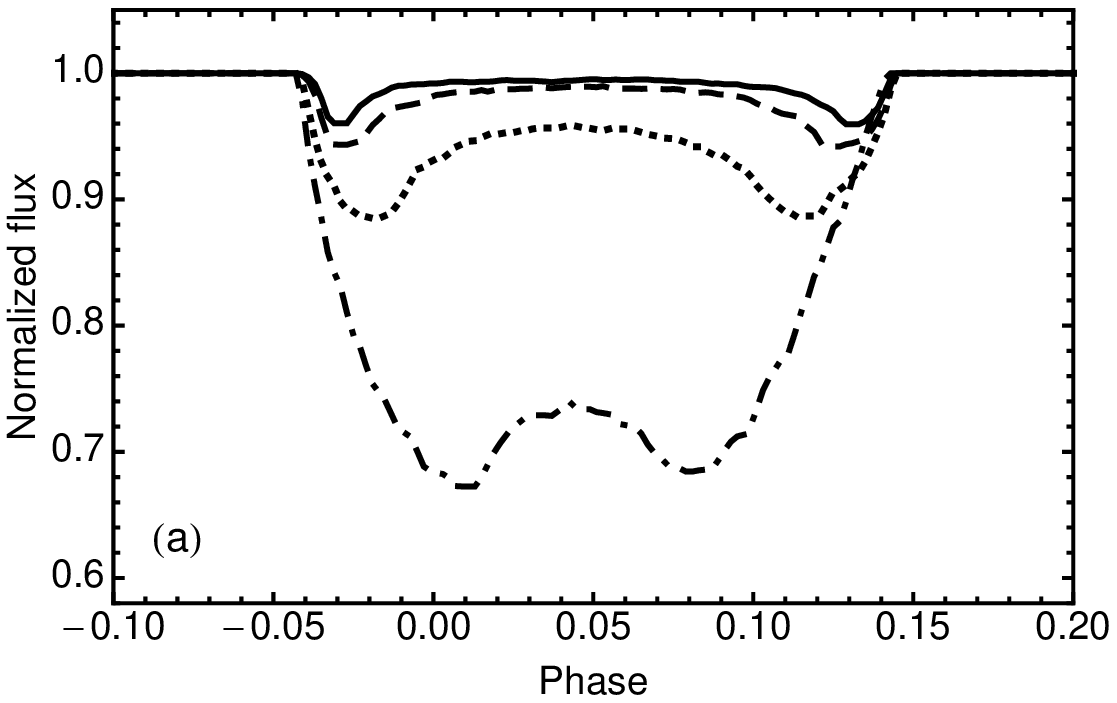}}\\
\centerline{\includegraphics[width=74mm]{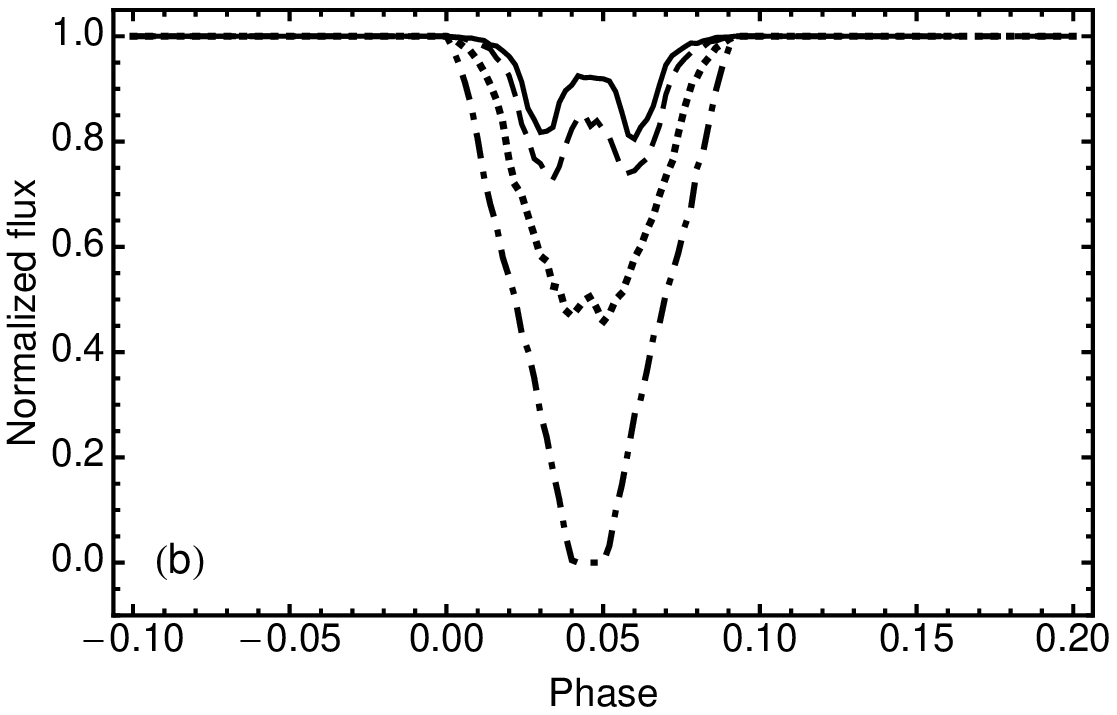}}\\
\end{tabular}
\caption{
Lightcurves showing the effect of occultation in LS 5039
(a) $i=60^{\circ}$ and $\eta_0 = 0.6$ and (b) $i=90^{\circ}$ and $\eta_0 = 0.004$ (measured at periastron). The curves correspond to emitting regions with sizes $R_{\rm out}$, $15 R_{\star}$ (solid), $10 R_{\star}$ (dashed), $5 R_{\star}$ (dotted), $2 R_{\star}$ (dot-dashed).}
\label{lightcurves_occultation}
\end{figure}

%%%%%%%%%%%%%%%%%%%%%%%%%%%%%%%%%%%%%%%%%%%%%%

\section{Occultation in LS 5039}
\label{s:occultation}
The occultation of the X-ray emitting region by the massive star does not depend on energy, thus its only signature in the spectrum is a periodic reduction in flux. The amplitude and duration of occultation depend on three main parameters $i$, $\eta_0$ and $R_{\rm out}$. We find that the duration of the occultation is longest for binaries with circular orbits and where at all phases $\eta_0 > 1$ (when the CD curves around the massive star). The occultation duration can last up to 40\% of the orbital period. The latter is because curved portions of the CD are partly occulted by the star even at phases away from the superior conjunction. The amplitude of occultation is highest if the inclination is high and the size of the emitting region is small compared to the size of the star. However, this situation is unlikely to occur in LS 5039 since $\eta_0\la 0.6$ (see \S\ref{s:ls5039}). %
\begin{figure}
\centerline{\includegraphics[width=75mm]{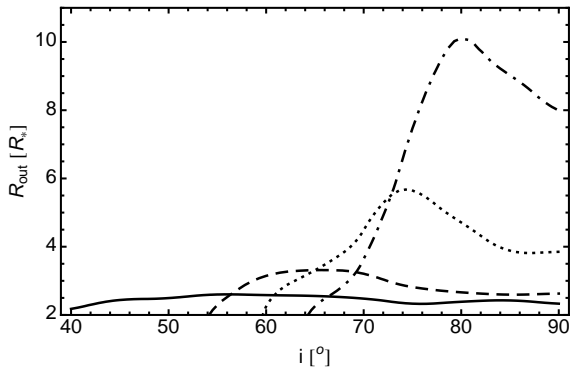}}
\caption{The areas above each curve correspond to permitted values of $i$ and $R_{\rm out}$ for which effect of occultation at superior conjunction does not exceed the $3\sigma_{\rm d}$ limit. Each curve corresponds to different value of $\eta_0$, solid 0.6, dashed 0.08, dotted 0.02, dot-dashed 0.004 measured at periastron.}
\label{occultation}
\end{figure}

When $\eta_0$ is below unity, occultation influences the shape of the X-ray lightcurve only around superior conjunction. Lightcurves calculated for the most preferable conditions to observe occultation in LS 5039 (high $\eta_0$, high $i$ and small $R_{\rm out}$) are shown in Fig. \ref{lightcurves_occultation}a. The dip in the lightcurve caused by occultation is narrow and covers only $\Delta \phi \sim 0.2$ in phase around superior conjunction. The depth of the minimum strongly depends on $R_{\rm out}$. Note also that for large $R_{\rm out}$ two minima appear, separated by a local maximum. This effect is related to the size of the shadow cast by the star onto the CD surface, which is larger when the line of sight is tangential to the surface of CD. These two local minima are too narrow ($\Delta \phi \sim 0.025$) to be resolved in the {\it Suzaku} lightcurve. A similar lightcurve study for $\eta_0 \ll 1$ shows even deeper and narrower minimum than in case of $\eta_0 \sim \eta_{\rm max}$ (Fig. \ref{lightcurves_occultation}b).

The X-ray modulation cannot be explained by occultation alone but this does not preclude an observable effect in the form of a sharp drop in flux around superior conjunction. In order to place an upper limit on this effect, we normalize the {\it Suzaku} lightcurve to 1 and fit it with a sine function in order to remove the orbital modulation. The standard deviation of the subtracted data is $\sigma_{\rm d}=0.1$. The upper limit on the depth of a dip in the data is assumed to be $3\sigma_{\rm d}$, meaning a $\leq30\%$ reduction of the 1--10 keV flux would not be statistically significant. The dip in the lightcurve at the superior conjunction may be caused by the combined effects of absorption and occultation. The effect of absorption on 1--10 keV flux is low. In a simple estimate, for $N_{\rm H}=N_{\rm H}^{\rm max}$ and for the observed power law spectrum of LS 5039 with photon index $\Gamma=1.51$, the integrated absorbed (using uniform absorber, i.e., wabs model in {\tt xspec}) 1--10 keV flux is at most $\sim 6\%$ lower than the 1--10 keV unabsorbed flux. Therefore we do not take absorption into account in the following calculations.

There is no significant dip in the data of LS 5039 at the phases close to superior conjunction. To put constraints on the binary parameters based on the lack  of occultation only, we explore the parameter space to find solution to equation in superior conjunction
\begin{equation}
{1 \over N} \sum_{j=1}^N  \zeta_j(i,\phi) = 1-3\sigma_{\rm d}.
\label{eq:maxoccultation}
\end{equation}

The final product of the study is Fig \ref{occultation} where areas above each curve correspond to inclination angles and sizes of the X-ray emitter for which the effect of occultation at superior conjunction does not exceed the $3\sigma_{\rm d}$ limit. The range of allowed parameters changes with the shape of CD.

\section{Conclusions}
\label{s:discussion}
Using a 3D model of a gamma-ray binary in the framework of the pulsar wind scenario, we tested the influence of X-ray absorption and occultation on the lightcurve and spectrum of LS 5039. We find that occultation cannot be responsible for the smooth X-ray orbital modulation in LS 5039, since
this would require $\eta_0>1$. This appear unlikely because there is no sign in the UV lines \citep{mcswain2004} that the O star wind is quenched on the hemisphere facing the compact object. Even for the most favorable conditions ($\eta_0=\eta_{\rm max}\approx0.6$ and large $i$) the duration of occultation never exceeds $\Delta\phi =0.2$. The cause of the X-ray modulation has to be found elsewhere \citep*{takahashi2009,dubus2010}.

Constraints on the binary geometry are derived from the observed upper limit on the intrinsic column density and on the depth of the occultation dip. Limiting the effects of absorption and occultation requires the X-ray source to be extended. The minimum size of the X-ray emitting region, which depends on the stellar mass loss rate, shape of CD and inclination angle, varies between 3--15 $R_{\star}$. An emitter size $> 4R_{\star}$ with $\eta_0 \ga 0.02 $ would be compatible with a $90^\circ$ binary inclination. The limit based on the lack of X-ray eclipses and assuming a point source at the compact object location is 60$^\circ$  \citep{casares2005}. A large X-ray source also loosens the constraints on the stellar mass loss rate derived from the observed lack of intrinsic absorption. For example, for $\eta_0 \sim 0.004$, an emission region $\sim 3 R_{\star}$ at inclination $\sim 50^{\circ}$, allows a mass-loss rate of $1.5 \times 10^{-7}$ M$_{\odot}$ yr$^{-1}$ compared to $4.7 \times 10^{-8}$ M$_{\odot}$ yr$^{-1}$ for  a point source located at the pulsar position. A more precise determination of the stellar wind mass loss rate would greatly help narrow down the possibilities. Still, we can conclude that an extended X-ray source appears necessary to reconcile the pulsar wind scenario with the upper limits on X-ray absorption in the wind and the absence of occultation features in the lightcurve. 

An extended X-ray source in gamma-ray binary can be expected in the pulsar wind scenario. High-energy electrons will be accelerated and randomized all along the termination shock of the pulsar wind. The shock distance from the pulsar is smaller towards the stellar companion than away from the O star, the efficiency of particle acceleration may also change at different locations. Calculations also show that the synchrotron emission from the electrons peaks in the 1--10 keV range only after significant cooling (\citealt{dubus2006}; \citealt*{dubus2008}). For a typical magnetic field of 0.1--1 G, the injected electrons radiate primarily synchrotron above 0.1 MeV while they upscatter stellar photons to energies above a GeV. The electrons are advected away in the shocked flow with an initial speed $\approx c/3$. The bulk of  the 1--10 keV radiation is emitted at a distance  $\approx 3 R_{\star}$ from the star in Fig.~ 4 of \citet{dubus2006}. More detailed modeling is required to obtain the exact evolution of the shock conditions with distance and quantify precisely the extent of the X-ray emission region. Such modeling would also yield the precise contribution to the X-ray emission and absorption from the shocked stellar wind region, which we have neglected here (\S\ref{s:model}).

The application to other known gamma-ray binaries is not straightforward since these contain a Be star with a dense equatorial outflow in addition to the tenuous stellar wind. The geometry of the interaction region with the pulsar wind changes between the polar wind and disk wind and can have a complicated shape at the transition that has yet to be investigated. However, the wider orbits of LS I+61$^\circ$303 and PSR B1259-63, the estimated inclination of $36^{\circ}$ in PSR B1259-63, will limit the impact of absorption and occultation on the lightcurve.

\section*{Acknowledgments}
This work was supported by the European Community via contract ERC-StG-200911 and in part by the Polish MNiSW grant NN203065933.

\bibliographystyle{mn2e} % style aa.bst
\bibliography{shock} % your references Yourfile.bib

\newpage

\section{Appendix A - Random generation of uniformly distributed points on the 3D surface.}

Here we explain how to generate a random sample of points uniformly distributed on a 3D surface $S$, symmetrical with respect to $x$-axis, described by a function $f(\rho)$ where $\rho=\sqrt{y^2+z^2}$.  We adopt an accept-reject algorithm commonly used for generating random samples drawn from an arbitrary distribution. 

First, we generate $N$ random points distributed uniformly on the $yz$-plane. The points are denoted in polar coordinates as $(\rho_1,\psi_1),(\rho_2,\psi_2),...,(\rho_{\rm N},\psi_{\rm N})$ where $\rho_{\rm i} \in \langle 0,\rho_{\rm max} \rangle$ and $\psi_{\rm i} \in \langle 0,2 \pi \rangle$. 

Second, we assign to each point a random number $w \in \langle 0,1\rangle$, where $w_1, w_2, ..., w_{\rm N}$ are drawn from an uniform distribution. The $w_{\rm i}$ are weights, used to accept and reject points on the basis of a density function
\begin{displaymath}
P(\rho_{\rm i}) \sim {dS \over M \rho_{\rm i} \Delta \rho \Delta \psi}  = 
\end{displaymath}
\begin{equation}
=  {1 \over M} \sqrt{1+ \left ({f(\rho_{\rm i}+\Delta \rho)-f(\rho_{\rm i}) \over \Delta \rho } \right  )^2 },\\
\end{equation}
where $\Delta\rho \rightarrow 0$, $P(\rho_{\rm i}) \in \langle 0, 1 \rangle$, $dS$ is a unit 3D surface element given by Eq.~\ref{eq:ds}, and
\begin{equation}
M= \sqrt{1+\left (  f(\rho_{\rm max}+\Delta \rho)-f(\rho_{\rm max}) \over \Delta \rho   \right )^2}.
\end{equation}

Finally, from our sample of points uniformly distributed on the 3D surface, we choose only points for which $w_{\rm i}  < P(\rho_{\rm i})$. In the paper, $f$ is the solution to Eq.~\ref{eq:antokhin}, while $\rho_{\rm max}$ is determined by $R_{\rm out}$ and has to be found numerically. An example sample of points is plotted in Fig. \ref{3Dshock_schematics}.

\end{document}